\documentclass[twoside]{ilcws07}
\usepackage[latin1]{inputenc}
\usepackage[dvips]{graphicx,epsfig,color}
\usepackage{psfrag,wrapfig,rotating}
\usepackage{amssymb,amsmath,array}

\usepackage{epsf}
\usepackage{amsfonts}
\usepackage{cite}
\usepackage[small]{caption2}
\usepackage{graphics}

\def\slashchar#1{\setbox0=\hbox{$#1$}
   \dimen0=\wd0
   \setbox1=\hbox{/} \dimen1=\wd1
   \ifdim\dimen0>\dimen1
      \rlap{\hbox to \dimen0{\hfil/\hfil}}
      #1
   \else
      \rlap{\hbox to \dimen1{\hfil$#1$\hfil}}
      /
   \fi}

\newcommand{\be}{\begin{equation}}
\newcommand{\ee}{\end{equation}}
\newcommand{\eq}{\end{equation}}
\newcommand{\rb}{\underline{r}}
\newcommand{\kb}{\underline{k}}

\begin{document}
\title{
BFKL resummation effects in exclusive production of rho meson pairs at the ILC} 
\author{M.~Segond$^1$\protect\footnote{ \, speaker}, L.~Szymanowski$^{2,3}$, S.~Wallon$^1$
\vspace{.3cm}\\
1- \, Universit\'e
Paris-Sud- LPT\footnote{\,\,Unit{\'e} mixte 8627 du CNRS} \\ 
 91405-Orsay- France 
\vspace{.1cm}\\
2-\,Universit\'e  de Li\`ege \\
  B4000  Li\`ege-Belgium
\vspace{.1cm}\\
3-\, So{\l}tan Institute for Nuclear Studies\\
Ho\.za 69, 00-681 Warsaw- Poland
}

\maketitle

\begin{abstract}
We calculate the Born order cross-section for the exclusive production
of rho meson pairs in $e^+e^-$ scattering in the Regge limit of QCD and we
show the feasibility of the measurement of this process at the ILC. 
We also investigate the leading and next-to-leading order BFKL evolution, making this process  a very clean test of the 
BFKL resummation effects.
\end{abstract}

\section{Impact factor representation in the Regge limit of QCD}

\label{sec:figures}

\psfrag{p1}[cc][cc]{$k_1$}
\psfrag{p2}[cc][cc]{$k_2$}
\psfrag{q1}[cc][cc]{$q_1$}
\psfrag{q2}[cc][cc]{$q_2$}
\psfrag{l1}[cc][cc]{}
\psfrag{l1p}[cc][cc]{}
\psfrag{l2}[cc][cc]{}
\psfrag{l2p}[cc][cc]{}
\psfrag{r}[cc][cc]{$r$}
\begin{wrapfigure}{r}{0.35\columnwidth}
\centerline{\includegraphics[width=0.3\columnwidth]{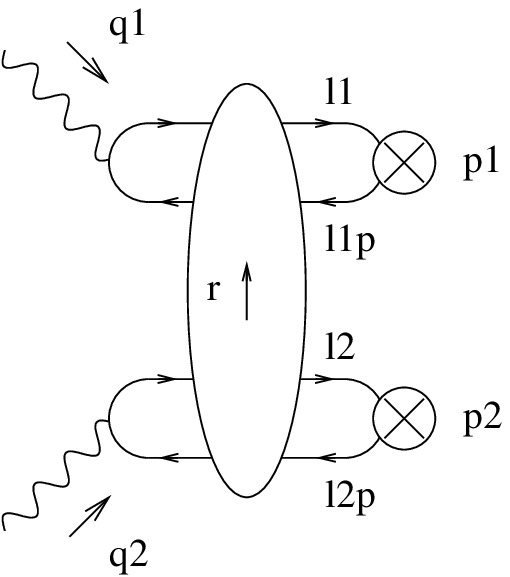}}
\caption{\small The amplitude of the process $\gamma^*_{L,T}(q_1) \gamma^*_{L,T}(q_2) \to \rho_L^0(k_1)  \rho_L^0(k_2)$
in the impact representation.\vspace{-0.50cm}}
\label{impact}
\end{wrapfigure}
We are focusing on the high-energy (Regge) limit, when the cm energy $s_{\gamma^*\gamma^*}$ is much larger than all other scales of the process, in which $t-$channel gluonic
exchanges dominate\cite{born}. 
The  highly virtual photons (the virtualities $Q^2_i=-q^2_i ,$ supply the hard scale which justifies the use of perturbation theory) provide small transverse size objects ($q \bar{q}$ color dipoles) whose scattering by pairs is the cleanest place to study the typical Regge behaviour with $t-$channel  BFKL Pomeron exchange\cite{bfkl}, in perturbative QCD. 
  If one selects the events with
comparable photon virtualities, the BFKL resummation effects dominate with respect to the
conventional partonic evolution of DGLAP\cite{dglap} type. Several studies of BFKL dynamics have been performed at the level of the total cross-section\cite{bfklinc}. At high energy, the impact factor representation of the scattering amplitude 
 has the form of a convolution in the transverse momentum  $\kb$ space between the  two  impact factors, corresponding  to the
transition of
$\gamma^*_{L,T}(q_i)\to \rho^0_L(k_i)$, 
via the $t-$channel exchange of two reggeized gluons (with momenta $\kb$ and $\rb -\kb$). The final states $\rho$ mesons are described in the collinear factorization by  their distribution amplitudes (DA) in a similar way as in the classical work of Brodsky-Lepage\cite{BLphysrev24}.



\section{Non-forward Born order cross-section at ILC for $e^+e^- \to e^+e^- \rho_L^0  \;\rho_L^0$ }

Our purpose is to evaluate at Born order and in the non-forward case the cross-section of the process $e^+e^- \to e^+e^- \rho_L^0  \;\rho_L^0$ in the planned  experimental conditions 
of the International Linear Collider (ILC). 

\begin{wrapfigure}{r}{0.68\columnwidth}
\vspace{-1.2cm}
\begin{center}
\begin{tabular}{lc} \hspace{-2.25cm}\epsfxsize=5.2cm{\centerline{\epsfbox{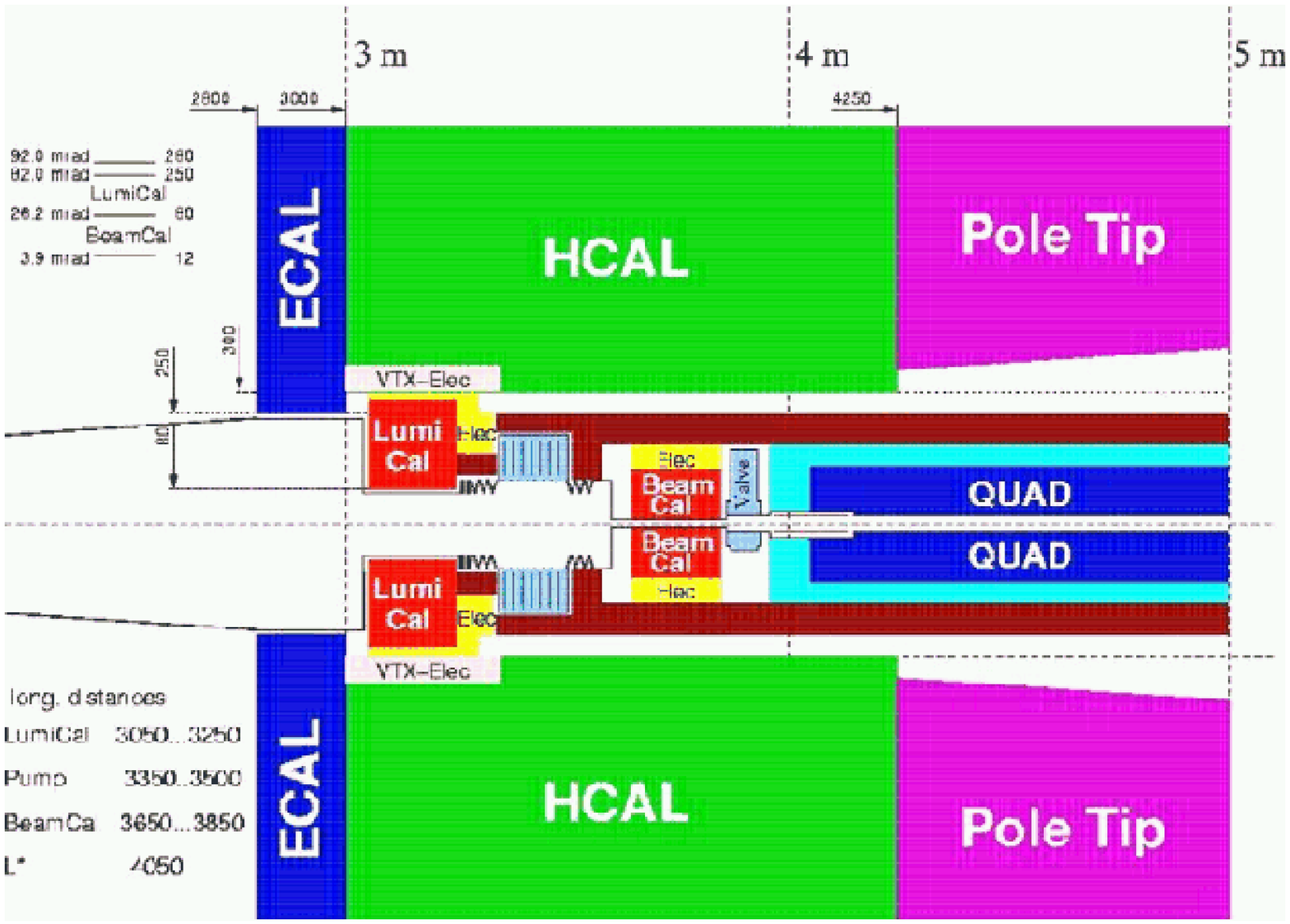}}} & \hspace{-5cm}\epsfxsize=3.65cm{\centerline{\epsfbox{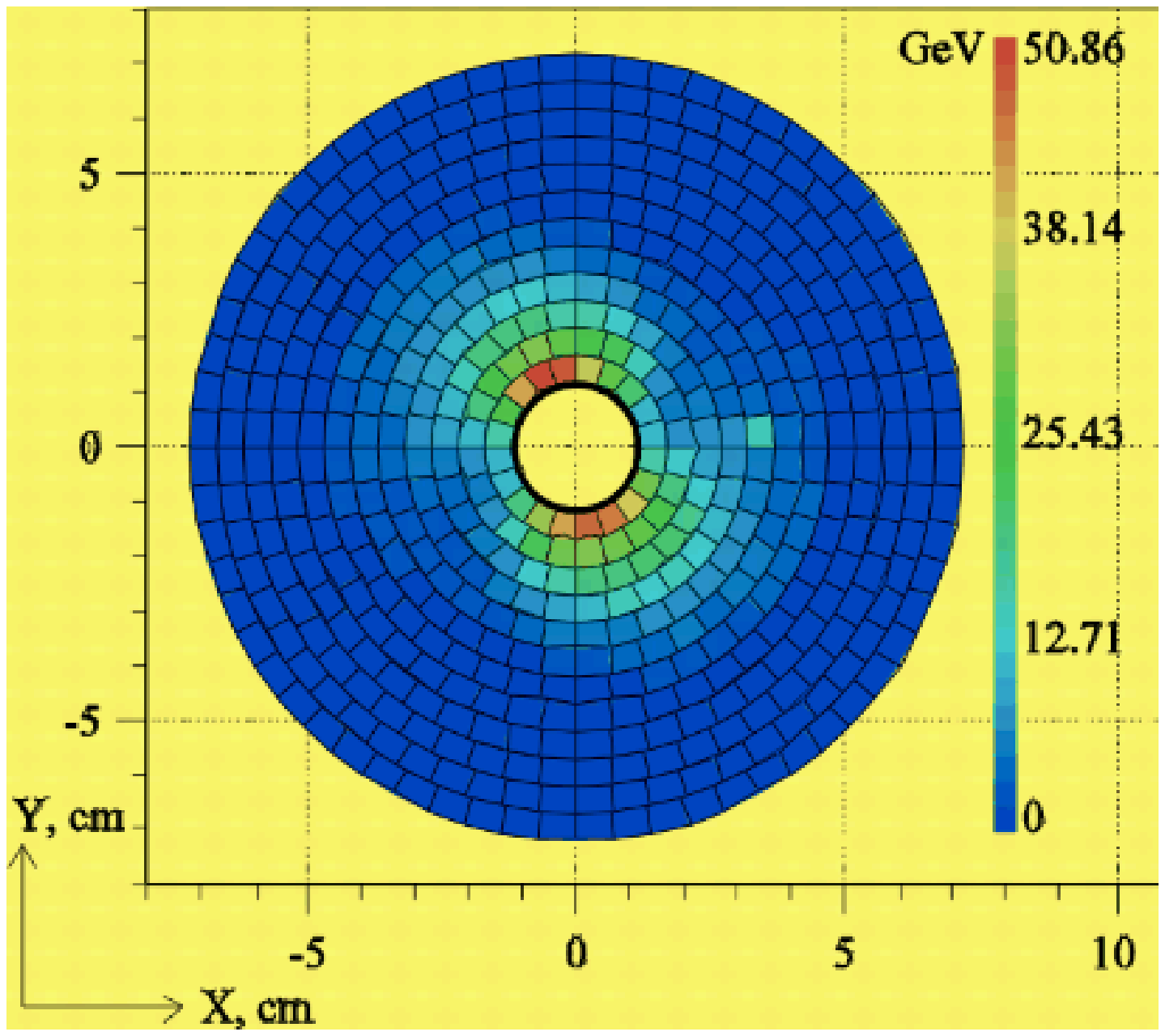}}}
\end{tabular}
\end{center}\vspace{-.5cm}
\caption{ \hspace{0.6cm}LDC  (a). \,   \hspace{1.2cm}Beamstrahlung in BeamCal (b).}
\label{FigLDC}\vspace{-.4cm}
\end{wrapfigure}

\noindent
We focus on the LDC detector project and we use the potential of the very forward
region accessible through the electromagnetic calorimeter  BeamCal which may be installed
around the beampipe at 3.65 m from the vertex. This calorimeter allows to detect (high energetic) particles down to 4 mrad. This important technological step was not feasible a few years ago. The foreseen energy of the collider is $ \sqrt{s}=500$ GeV.
Moreover we  impose that $s_{\gamma^*\gamma^*}  >c \, Q_1 \, Q_2$ (where $c$ is an arbitrary constant): it is required by the Regge kinematics for which the impact representation is valid. 
We choose $Q_{i\, min}=1$ GeV and $Q_{i\, max}=4$ GeV: indeed the various amplitudes
involved are completely negligible for higher values of virtualities.

\begin{wrapfigure}{r}{0.45\columnwidth}
\centerline{\includegraphics[width=0.4\columnwidth]{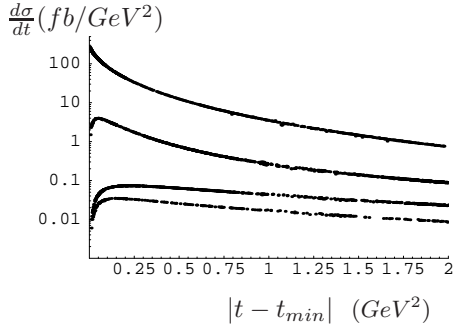}}
\begin{picture}(10,20)
\put(80,10){ $|t-t_{min}| $ \  \small $ (GeV^2)$}
\put(0,120){$\frac{d\sigma}{dt} (fb/GeV^2)$}
\end{picture}
\caption{\small Cross-sections for $e^+e^- \to e^+e^- \rho_L^0  \;\rho_L^0$ process. Starting from above, we display the
 cross-sections  corresponding to the  $\gamma^*_L \gamma^*_L$ mode,  to the $\gamma^*_L\gamma^*_T$ modes,   to the 
$\gamma^*_T \gamma^*_{T'}$ modes with  different $T \neq T'$ and finally to the $\gamma^*_T \gamma^*_{T'}$ modes with $T=T'$.}
\label{FigLogcurves}
\end{wrapfigure}

\begin{wrapfigure}{r}{0.5\columnwidth}
\vspace{-7.65cm}
\centerline{\includegraphics[width=0.4\columnwidth]{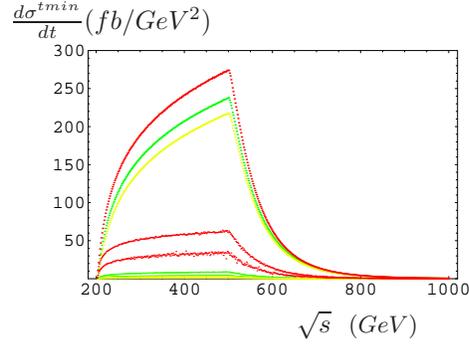}}
\begin{picture}(10,10)
\put(110,0){ $\sqrt{s} $ \  \small $ (GeV)$}
\put(5,115){$\frac{d\sigma^{tmin}}{dt} (fb/GeV^2)$}
\end{picture}
\caption{\small Cross-sections for $e^+e^- \to e^+e^- \rho_L^0  \;\rho_L^0$ at $t=t_{min}$ for different values of the parameter $c$: the red curves correspond to  $c=1$, the green curves to $c=2$ and and the yellow curves to $c=3$. For each value of c, by decreasing order the curves correspond to gluon-exchange, quark-exchange with $\gamma^*_L$ and quark-exchange with $\gamma^*_T$ .\vspace{-0.2cm} }
\label{Figceffects}
\end{wrapfigure}

\vspace{0.1cm}
We now display in Fig.\ref{FigLogcurves} the cross-sections  as a function of the momentum transfer $t$ for the different $\gamma^{*}$ polarizations. For that we performed analytically the integrations over  $\kb$ (using conformal transformations to reduce the number of massless propagators) and numerically the integration over the accessible phase space.
 We assume the QCD coupling constant to be $\alpha_{s}(\sqrt{Q_1 Q_2})$ running at three loops, the parameter $c=1$ which enters in the Regge limit condition and the 
 energy of the beam $ \sqrt{s}=500$ GeV. We see that all the differential cross-sections which involve at least one transverse
photon vanish in the forward case when $t=t_{min}$, due to the $s$-channel helicity conservation.
We finally display in the Table.\ref{tab} the results for the total cross-section integrated over $t$ for various values of c.  With the foreseen nominal integrated luminosity of $125  \, {\rm fb}^{-1},$ this will yield $4.26\, 10^3$ events per year with $c=1$.

\vspace{-0.4cm}
\begin{wraptable}{l}{0.2\columnwidth}
\centerline{\begin{tabular}{|l|r|}
\hline
c  & $\sigma^{Total} \, ( fb )$ \\\hline  
1  & 34.1 \,  \,    \\\hline
2         & 29.6  \,  \, \\\hline
10             & 20.3  \,  \,    \\\hline
\hline
\end{tabular}}
\vspace{-0.2cm}
\caption{ $\sigma^{Total}$ for various c.\vspace{-0.4cm}}
\label{tab}
\end{wraptable}

\hspace{-1.1cm}
By looking into the upper curve in the Fig.\ref{FigLogcurves} related to the longitudinal polarizations, one sees that the point  $t=t_{min}$ gives the maximum of the total cross-section (since the transverse polarization case vanishes at $t_{min}$) and then practically dictates the trend of the total cross-section which is strongly peaked in the forward direction (for the longitudinal case) and strongly decreases with $t$ (for all polarizations).
From now we only consider the forward dynamics. The Fig.\ref{Figceffects} shows the cross-section (for both gluons and quarks exchanges) at $t_{min}$ for different values of the parameter c which enters in the Regge limit condition: the increase of c leads to the suppression of quarks exchanges (studied in\cite{gdatda}) and supplies us an argument to fix the value of $c$ on the gluon exchange dominance over the quark exchange contribution. 
The ILC collider is expected to run at a cm nominal energy of 500 GeV, though it might be extended in order to cover a range between 200 GeV and 1 TeV.  Although the Born order cross-sections do not depend on $s,$ the triggering effects introduce an $s$-dependence that explains the peculiar ('fin of shark' like) shape of the cross sections displayed in Fig.\ref{Figceffects}: because we chose $Q_{i\, min}=1$GeV  (as hard cut required by  the perturbative analysis), the corresponding minimal angle of the scattered leptons (that behaves like $2 \, Q_{i\, min}/ \sqrt{s} $) will cross over the experimental cut imposed by the resolution of the calorimeter as soon as  $\sqrt{s}$ will be bigger than 500 GeV, explaining why the cross-sections fall down between 500 GeV and 1 TeV. The measurability is then optimal for $\sqrt{s}=500$ GeV.

\section{Forward differential cross-section with BFKL evolution}

\begin{wrapfigure}{r}{0.5\columnwidth}
\centerline{\includegraphics[width=0.4\columnwidth]{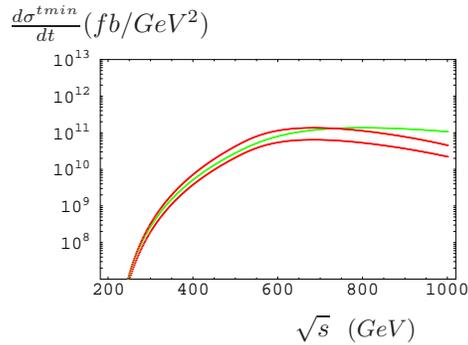}}
\begin{picture}(10,10)
\put(110,0){ $\sqrt{s} $ \  \small $ (GeV)$}
\put(5,115){$\frac{d\sigma^{tmin}}{dt} (fb/GeV^2)$}
\end{picture}
\caption{\small Cross-sections at $t=t_{min}$ for $e^+e^- \to e^+e^- \rho_L^0  \;\rho_L^0$ with LO BFKL evolution for different $\alpha_s$ : the upper and lower red curves for  $\alpha_s$ running respectively at one and three loops  and the green one for $\alpha_s = 0.46$.  }
\label{Figbfkltmin}
\end{wrapfigure}

\begin{wrapfigure}{r}{0.45\columnwidth}
\vspace{-6.35cm}
\centerline{\includegraphics[width=0.4\columnwidth]{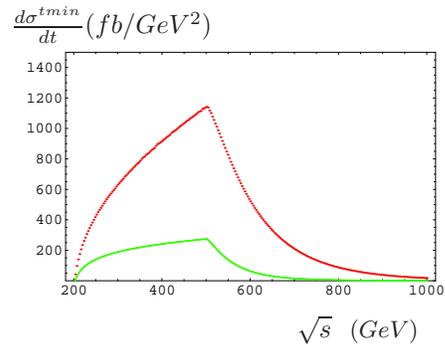}}
\begin{picture}(10,10)
\put(110,0){ $\sqrt{s} $ \  \small $ (GeV)$}
\put(5,115){$\frac{d\sigma^{tmin}}{dt} (fb/GeV^2)$}
\end{picture}
\caption{\small Cross-sections at $t=t_{min}$ for $e^+e^- \to e^+e^- \rho_L^0  \;\rho_L^0$ with NLO BFKL evolution   (red curve) and at Born order (green curve) .}
\label{FigNLOtmin}
\end{wrapfigure}

The results obtained at Born approximation can be considered as a lower limit of the cross-sections for $\rho$-mesons pairs production with complete BFKL evolution taken into account. We also consider below only the forward case and we first evaluate  the leading order (LO) BFKL evolution (in the saddle point approximation) of our process. The comparison of Fig.\ref{Figceffects} with  Fig.\ref{Figbfkltmin} leads to the conclusion that the LO BFKL evolution dramatically enhances  the shape of the cross-section when increasing $\sqrt{s}$, though it is not very fruitful to make precise predictions: indeed, the Pomeron intercept (which corresponds to the leading pole in the $\omega$ plane and then controls the power like growth of the amplitude) takes quite large values mainly because it is proportional to the strong coupling $\alpha_{s}(\sqrt{Q_1 Q_2})$ whose scale dependence is arbitrarily prescribed at LO, and causes severe instabilities (since its running starts at 1GeV with our choice of the hard cut). It is well-known that the next-to-leading order (NLO) contribution is expected to be between the LO and Born order cross-sections since the value of the intercept is widely reduced when considering NLO BFKL evolution. To study these effects we use the renormalization group improved BFKL kernel\cite{NLLpaper}. Our results are in accordance with the ones made from the full NLO kernel used in\cite{papa}. In this approach developped in\cite{epsw}, we must find the solutions (the NLL Pomeron intercept $\omega_s$ and the anomalous dimension $\gamma_s$) of a set of two coupled equations.
Although this approach needs a fixed strong coupling, we reconstruct in $\omega_s$ and $\gamma_s$ a scale dependence by fitting with polynomials of  $Q_i$  a large range of solutions obtained for various values of $\alpha_s(\sqrt{Q_1 Q_2})$. Moreover, the results are much less sensitive to the choice of the scale dependence of the strong coupling than the ones obtained at LO. We display in the Fig.\ref{FigNLOtmin} the curves (with $c=1$) at Born order obtained previously with the one obtained after NLO  BFKL resummation.


\vspace{0.161cm}

\noindent


\section{Acknowledgments}
We thank the denseQCD ANR for support and the organizers of the LCWS/ILC07 conference. L.Sz. thanks the support of Polish Grant 1 P03B 028 28. L.Sz. is a visiting fellow of FNRS (Belgium).
\section{Bibliography}

\begin{footnotesize}



%

\end{footnotesize}


\end{document}